\documentclass[final,12pt]{elsarticle} 

\usepackage{graphicx} 
\usepackage{tabularx} 
\usepackage{epstopdf} 
\usepackage{dcolumn} 
\usepackage{xcolor} 
\usepackage{amsmath,amssymb} 
\usepackage[hidelinks]{hyperref} 
\usepackage[utf8]{inputenc} 
\usepackage[english]{babel} 
\usepackage[displaymath, mathlines]{lineno} 
\usepackage{blindtext} 
\usepackage{ifthen} 
\usepackage{orcidlink} 
\usepackage{multirow}
\usepackage[section]{placeins}
\usepackage{subcaption}
\usepackage{float}
\usepackage{dblfloatfix}
\usepackage{setspace}
\usepackage{ulem} 

\newboolean{articletitles}
\setboolean{articletitles}{True} 

\usepackage{environ}
\NewEnviron{comment}{}
\newcommand\con{\RenewEnviron{comment}{\color{olive}\BODY}}

\newcommand{\cmt}[1]{\begin{comment}{#1}\end{comment}\xspace}
\con
\usepackage[firstpage]{draftwatermark}

\newcommand{\changecolor}{blue}

\newcommand{\changeoff}{\renewcommand{\changecolor}{black}}

\changeoff

\newboolean{showstrikethrough}

\makeatletter
\newcommand{\conditionalStrike}[1]{%
    \ifthenelse{\boolean{showstrikethrough}}%
        {\textcolor{red}{\sout{#1}}}
        {%
            \@ifnextchar\bgroup{\ignorespaces}{
                \@ifnextchar\unskip\space{\ignorespaces}
            }%
        }%
}
\makeatother

\setboolean{showstrikethrough}{false}

\newcolumntype{Z}{>{\centering\let\newline\\\arraybackslash\hspace{0pt}}X}

\input{belle2-symbols}

\journal{Nuclear Physics A}
\onecolumn 

\begin{document}

\begin{frontmatter}

\title{\centering
{\large \textbf{ Development and commissioning of a new readout system for the gas flow of the Belle II \KL and muon detector}}\\ 
 \today}

\author[1]{T.~Applegate \orcidlink{0009-0000-2440-4999}}
\author[1]{N.~Brenny \orcidlink{0009-0006-2917-9173}}
\author[1]{C.~Chen \orcidlink{0000-0003-1589-9955}}
\author[1]{S.~Choudhury \orcidlink{0000-0001-9841-0216}}
\author[1]{J.~Cochran \orcidlink{0000-0002-1492-914X}}
\author[1]{S.~Kang \orcidlink{0000-0002-5320-7043}}
\author[1]{A.~Khatri \orcidlink{0009-0001-2662-3957}}
\author[2,3]{H.~Kindo \orcidlink{0000-0002-6756-3591}}
\author[2]{T.~Lam \orcidlink{0000-0001-9128-6806}}
\author[1]{S.~Mitra \orcidlink{0000-0002-1118-6344}}
\author[1]{A.~Mubarak \orcidlink{0000-0002-3529-4438}}
\author[2]{L.~Piilonen \orcidlink{0000-0001-6836-0748}}
\author[1]{S.~Prell \orcidlink{0000-0002-0195-8005}}
\author[1]{M.~Veronesi \orcidlink{0000-0002-1916-3884}}

\affiliation[1]{organization={Iowa State University},
        city={Ames},
        country={USA}}
\affiliation[2]{organization={Virginia Tech},
        city={Blacksburg},
        country={USA}}
\affiliation[3]{organization={High Energy Accelerator Research Organization},
        city={Tsukuba},
        country={Japan}}

\begin{abstract}
We have designed and commissioned a new readout board to detect photosensor signals from gas-bubbler panels to continuously monitor the gas flow through the resistive plate chambers of the \KL and muon detector of \belletwo.
The gas flow measurements have been integrated into \belletwo's alarm system.
The bubbler-monitoring system was first employed during the February 2024 to July 2024 \belletwo data-taking period.
\end{abstract}


\end{frontmatter}

\section{Introduction}
\label{sec:intro}
\begin{figure*}[htb!]
    \centering
    \includegraphics[width=0.9\linewidth]{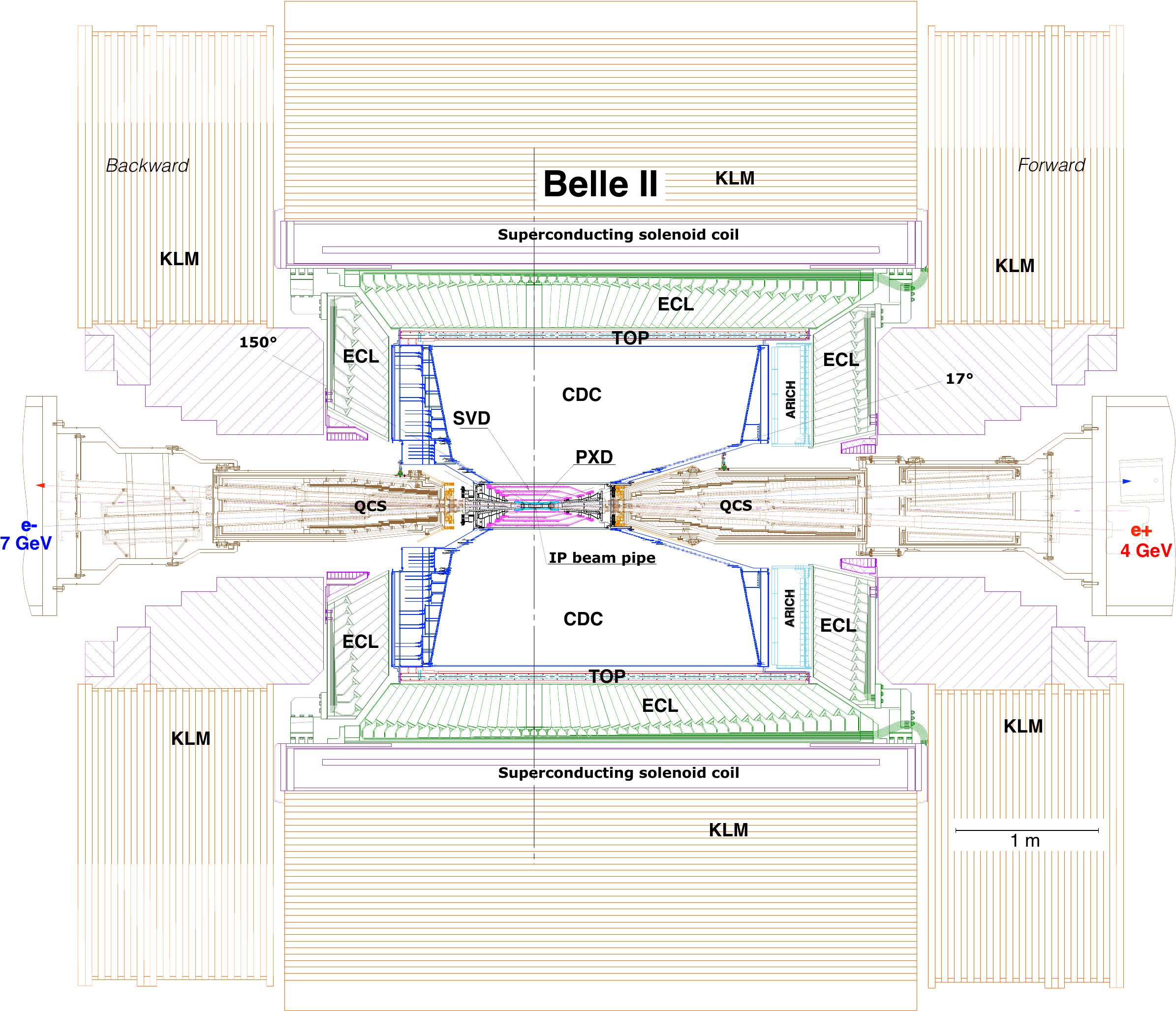}
    \caption{Cross-section view of the \belletwo detector from the top.}
    \label{fig:belleii-detector}
\end{figure*}

\begin{figure*}[htb!]
    \centering
    \begin{subfigure}[t]{0.45\textwidth}
        \centering
        \includegraphics[width=0.9\linewidth]{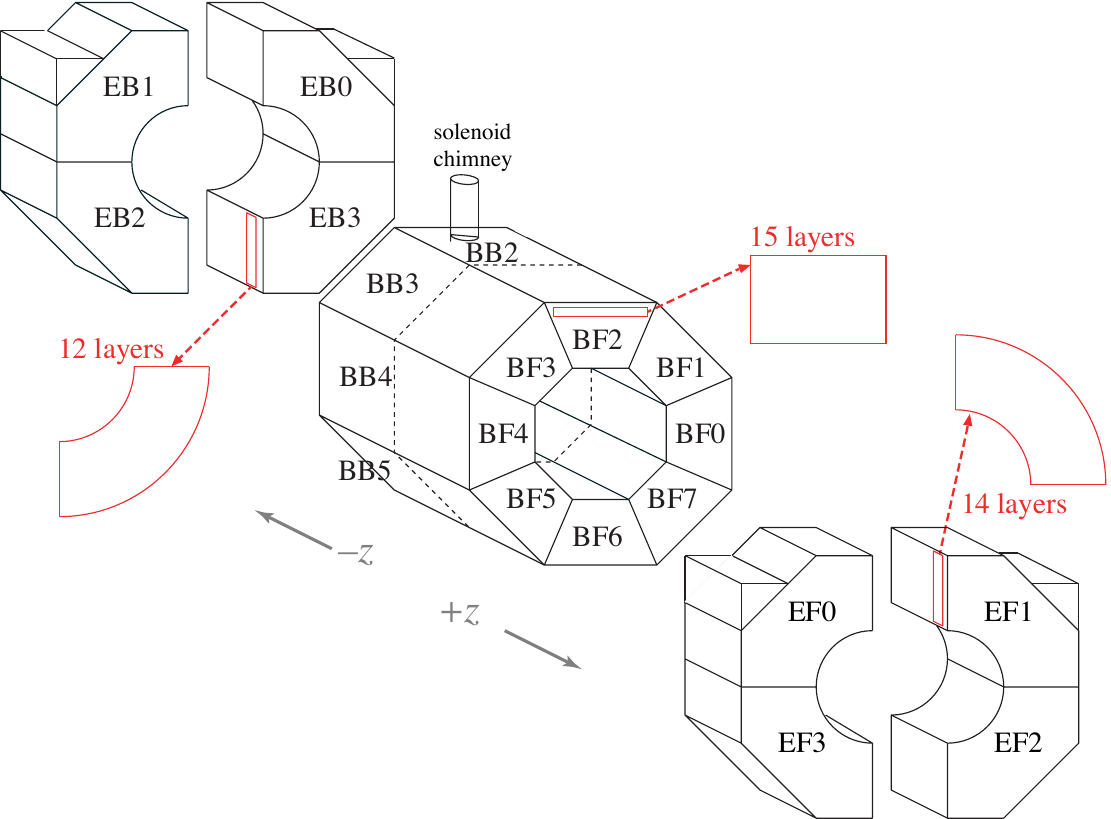}
        \caption{}
    \end{subfigure}
    \begin{subfigure}[t]{0.45\textwidth}
        \centering
        \includegraphics[width=0.9\linewidth]{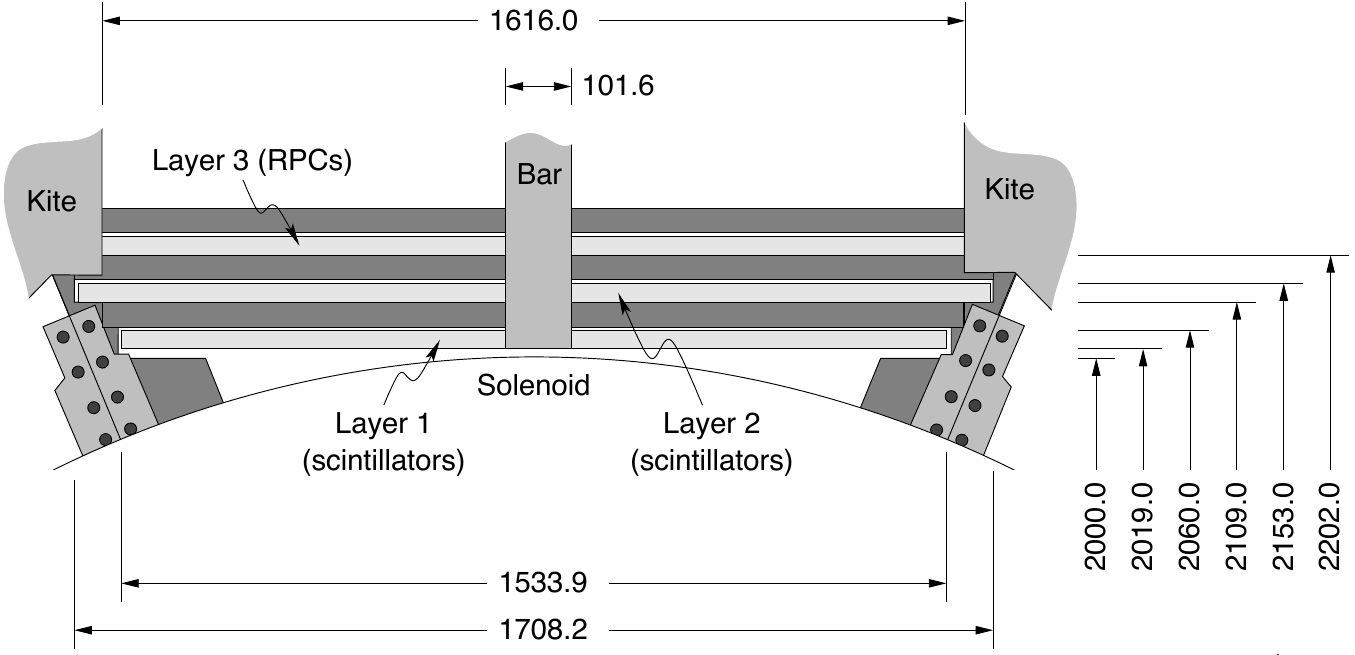}
        \caption{}
    \end{subfigure}
    \caption{Figure~(a) shows an exploded isometric view of the KLM subdetector. The $+z$ direction indicates the forward direction, while the $-z$ direction indicates the backward direction. Figure~(b) shows an end view of the first few layers of barrel octant BF2.}
    \label{fig:klm-detector}
\end{figure*}

The \KL and muon subdetector (KLM) of the \belletwo $B$-Factory experiment~\cite{10.1093/ptep/ptz106} operating at the SuperKEKB \epem collider identifies copiously produced \KL mesons and muons, which are useful for studying precision Standard Model and beyond the Standard Model physics. A cross-section view of the \belletwo detector from the top is shown in Fig.~\ref{fig:belleii-detector}.
The KLM is composed of three sections: a barrel and two endcaps (see Fig.~\ref{fig:klm-detector}~(a)). The barrel has a forward section facing the higher-momentum electron beam and a backward section facing the positron beam. Each barrel section has an octagonal structure subdivided symmetrically into eight sectors. 
In the barrel, the detector panels of the outer 13 layers are composed of glass-electrode resistive plate chambers (RPCs)~\cite{Abashian:2000vb}, while the inner two layers are composed of scintillating strips (see Fig.~\ref{fig:klm-detector}~(b)).
Each RPC panel is composed of an inner and outer RPC, which are independently operated. 
Each RPC in the panel has two parallel sheets of float glass for electrodes, separated by noryl spacers, epoxied in place.
All layers in the endcaps use scintillating strips.
Three-dimensional measurements of a particle's position are obtained from hits on orthogonal readout strips in each layer.
The RPCs, including the gas-flow measuring bubblers, were installed in the \belle detector \cite{Belle:2000cnh}, \belletwo's predecessor, and were in use from 1999 to 2011.
The bubbler readout  electronics, however,  were not included in the initial upgrade to \belletwo.
See Tab.~\ref{tab:upgrade-timeline} for a timeline of the KLM upgrade from Belle to \belletwo.
\begin{table}[htb!]
    \centering
    \begin{tabularx}{\textwidth}{|c|Z|}
    \hline
    \textbf{Date} & \textbf{Upgrade} \\
    \hline
    $2013-2014$ & Replace RPCs in the two inner barrel layers and in all of the endcap layers with scintillators \\
    \hline
    $01/2016-06/2016$ & Install scintillator readout electronics \\
    \hline
    $01/2017-06/2017$ & Remove Belle-era readout electronics, including bubbler readout boards, during replacement of RPC readout electronics \\
    \hline
    \end{tabularx}
    \caption{Timeline of the KLM upgrade from Belle to \belletwo}
    \label{tab:upgrade-timeline}
\end{table}

It was discovered in April 2021 that a sector's RPC gas flow regulator had been accidentally switched off, and the affected RPCs had gone without gas flow for two years. Further investigations showed that one of the sixteen KLM barrel half-octants experienced a 50\% reduction in hit detection efficiency. The suspected culprit is moisture accumulation in the RPC gas.

To prevent another case of detector damage and detection efficiency loss, we have developed and installed a new readout and monitoring board that uses the existing Belle-era bubbler hardware.  In this article, we discuss the development and commissioning of the new readout board, which is composed of analog multiplexers, analog-to-digital converters, and a single board computer, to process the analog signals from the existing Belle-era bubbler panel readout boards. The new readout boards continuously monitor the KLM RPC gas flow by measuring the exhaust-line gas bubble rates received from the bubblers. The bubbler-monitoring system is integrated into the \belletwo alarm and archival system.

\section{Measuring the RPC gas flow}
\label{sec:measuring}
\subsection{Bubbler panels}
\label{subsec:bubblers}

\begin{figure}[htb!]
    \centering
    \includegraphics[width=1\linewidth]{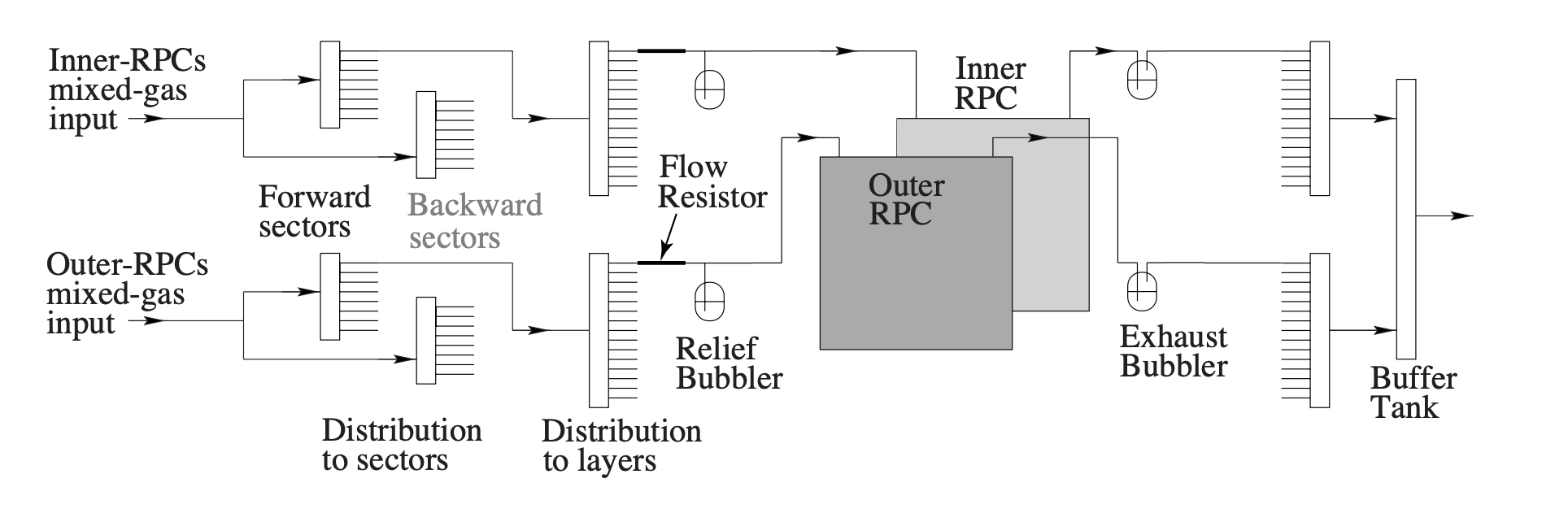}
    \caption{\belletwo RPC gas-distribution system.}
    \label{fig:gas-distribution}
\end{figure}

The RPCs in the barrel KLM operate with a gas mixture of 62\% HFC-134a, 30\% argon, and 8\% butane-silver.
During normal beam operation, the nominal flow rate is 2 l/min.
Fig.~\ref{fig:gas-distribution} shows the RPC gas-distribution system, which controls the injection of the gas mixture into distributors that direct the gas through the individual RPCs.
Between a distributor and each layer's RPC, there is a relief bubbler through which the gas can vent in case of overpressure.
After passing through the RPC, the gas flows through an exhaust bubbler and into a buffer tank.
The gas flow-rate through each RPC can be monitored by measuring the bubble rate in its exhaust bubblers.

\begin{figure}[htb!]
    \centering
    \includegraphics[width=1\linewidth]{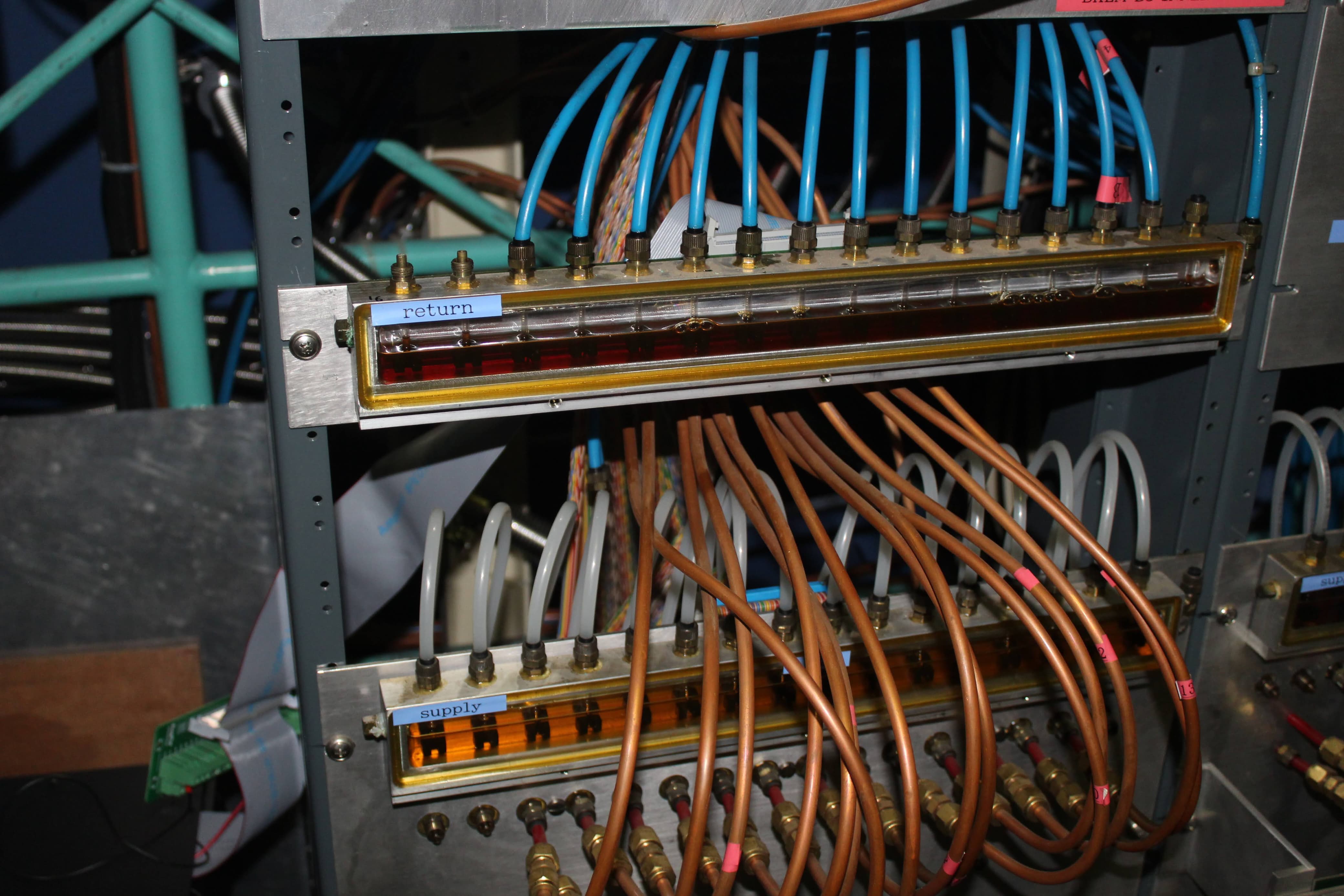}
    \caption{Two panels of 16 bubblers each, 13 of which are attached to RPC gas lines. The top panel has exhaust bubblers, and the bottom panel has relief bubblers.}
    \label{fig:panel}
\end{figure}

During normal operation, bubbles are expected in the exhaust bubblers but not the relief bubblers.
Flow in a relief bubbler indicates an overpressure condition that may damage the RPCs.
The exhaust and relief bubblers for a KLM sector are assembled on rack-mounted panels (see Fig.~\ref{fig:panel}).

Each bubbler is comprised of an open-ended tube for gas flow, an infrared LED, and a photosensor.
The photosensor is a silicon photodarlington transistor, and the LED-photosensor pair is the QT H22A1.
The cavity that the gas flows into is partially filled with mineral oil.
When the gas exits the tube into the oil, it creates a bubble.
 The bubble varies the amount of light from the LED that reaches the photosensor and thereby generates an analog signal.
The analog signals for the 13 bubblers for each sector are collected by a circuit that was deployed in the \belle era.
The circuit collects the analog signals and transmits them via a ribbon cable connector to our new readout boards.
(The Belle bubbler-readout  boards were removed during the KLM upgrade from Belle to \belletwo.)
In total, there are 832 bubbler rates to measure in the \belletwo KLM gas system: (1 exhaust + 1 relief) bubblers $\times$  2 RPCs/layer $\times$ 13 layers/octant $\times$ 8 octants/section $\times$ 2 forward/backward sections.

\subsection{Bubble rate measurement}
\label{bubble-rate-measurement}

\begin{figure*}[htb!]
    \centering
    \begin{subfigure}[t]{0.45\textwidth}
        \centering
        \includegraphics[width=1\linewidth]{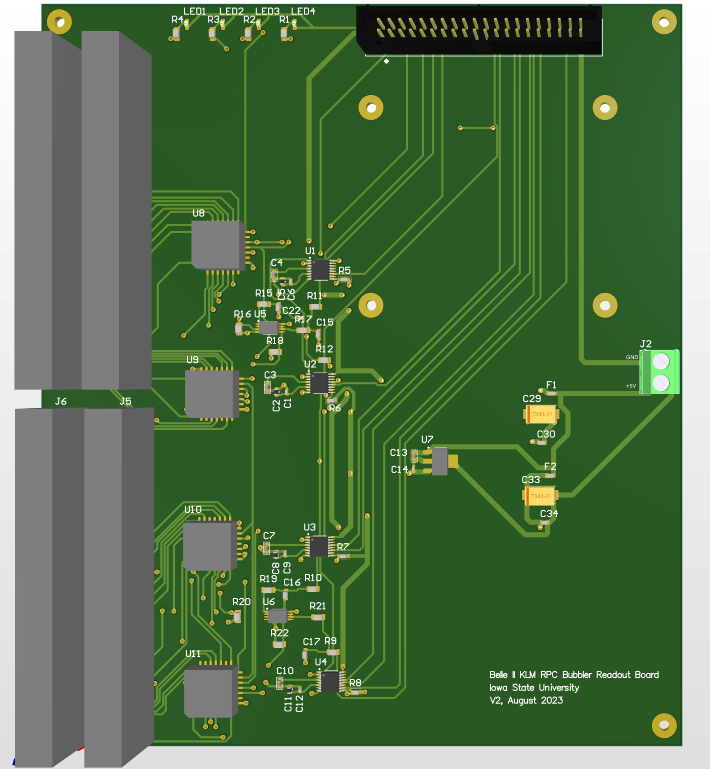}
        \caption{}
    \end{subfigure}
    \begin{subfigure}[t]{0.35\textwidth}
        \centering
        \includegraphics[width=1\linewidth]{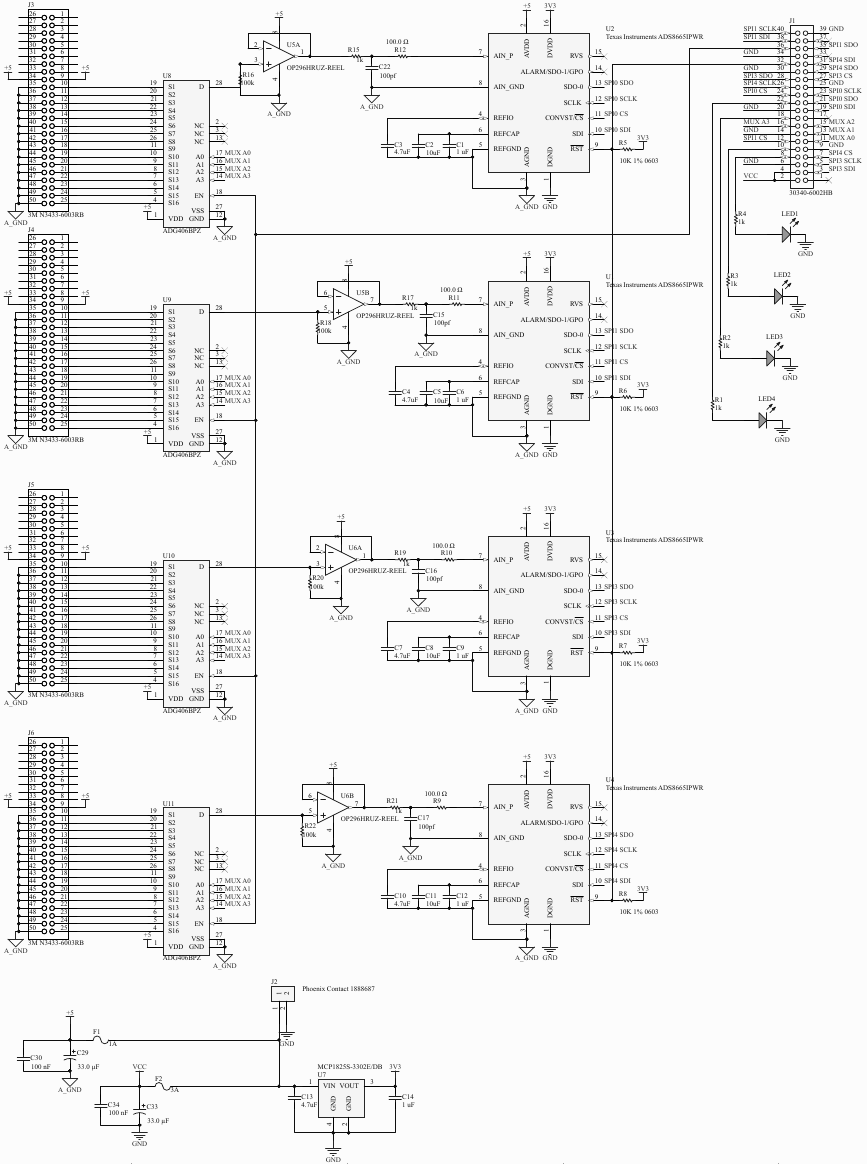}
        \caption{}
    \end{subfigure}
    \caption{Board design to digitize and multiplex the analog signals from the bubbler photodarlington transistors. Figure (a) shows the actual board, and figure (b) shows the circuit schematic. The SBC (not shown) is mounted near the pins at the top of the figure (a) and at label J1 in figure (b). The ribbon cables are attached at the gray ports on the left side of the figure (a). The main components are the multiplexers with labels U8-U11 and the ADCs with labels U1-U4.}
    \label{fig:circuit}
\end{figure*}

The analog signals from the photosensor are routed via ribbon cable to a newly designed printed circuit board.
We have designed the board to process analog signals from up to 4 $\times$ 16 channels (see Fig.~\ref{fig:circuit}).
The analog signals first pass through the board's 16-channel multiplexer, the ADG406.
Next, the signals are digitized by the board's 12-bit Analog-to-Digital Converter (ADC), the ADS8665. Each board has four ADCs to accommodate an entire sector consisting of exhaust and relief bubblers for each of the two layers of 13 RPCs.
The digital output from the ADCs is communicated to a board-mounted single-board computer (SBC), a Raspberry Pi 4 Model B, using the Serial Peripheral Interface protocol via the SBC's input/output pins.
The SBC also controls the multiplexer switches.
One entire process cycle, which consists of switching through 16 channels and for each channel collecting the digital output for each of the four ADCs, takes 2.4 milliseconds. The system can collect 250,000 sensor readings over a period of 10 minutes for each of the 4 ADCs $\times$ 16 channels/ADC = 64 channels.
A single bubble has a duration of 150-250 milliseconds and is accurately resolved with approximately 60 readings.

\begin{figure}[htb!]
    \centering
    \includegraphics[width=1\linewidth]{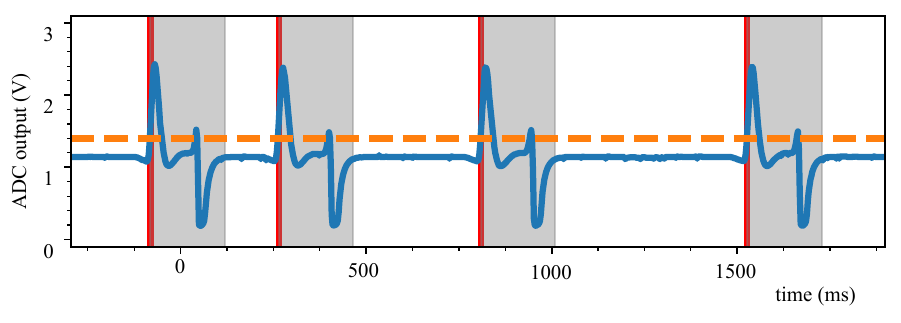}
    \caption{Characteristic voltage pulses from a bubble refracting infrared light from the photosensor. The horizontal axis shows time in milliseconds. The solid blue line is the voltage of the bubbler's photosensor. The dashed orange line is the voltage threshold used in the signal identification software. The red line indicates that a bubble has been triggered by the software, and the gray region following it is the minimum bubble width to avoid double counting. The figure shows a bubbling rate that is greater than the nominal integrated bubbling rate. The integrated bubbling rate averages out short periods of low and high bubbling rates throughout the integration period.}
    \label{fig:bubble-pulse}
\end{figure}

A bubble's passage in front of an LED refracts the light away from the photosensor, creating a characteristic voltage pulse (see Fig.~\ref{fig:bubble-pulse}).
The software on the SBC can distinguish the signal pulses from the otherwise constant voltage.
The signal-recognition software triggers on a rising edge of a pulse through a voltage threshold as well as a minimum bubble width threshold to avoid double-counting.
The bubbling rates are calculated periodically by integrating the number of detected bubbles during a fixed time interval.
Typical bubbling rates for exhaust bubblers are $0.1 - 0.5$ Hz with standard deviation $0.01-0.04$ Hz.

Each board processes analog signals from 4 ribbon cables $\times$ 13 channels = 52 channels using four multiplexers, four ADCs, and one SBC.
We have packaged four boards into a crate together with a power supply and an Ethernet switch.
The power supply in the crate is powered by a network-connected 110 V Power Distribution Unit, which allows for remote power cycling.
In total, we have built four crates containing four boards each.

\section{Integration with \belletwo}
\label{sec:monitoring}
\subsection{Network communication}
Each crate is connected via Ethernet to the \belletwo data acquisition network, Network Shared Memory (b2nsm)~\cite{b2nsm}, which enables network communication with the SBCs through the crate's Ethernet switch.
Bubble rates are recorded and stored by \belletwo's EPICS-based archiving system~\cite{Kim_2021} every $10$ minutes.
 They are communicated as EPICS Process Variables (PVs) via b2nsm to operators in the \belletwo control room.
Abnormal flow rates trigger an alarm.

\subsection{Alarm configuration}
\label{subsec:alarm-config}
Each bubble rate PV has an associated alarm status, which is integrated into the \belletwo detector control and alarm system~\cite{alarm}.
The alarm criteria have been configured to be sensitive to interruptions in individual RPC bubbling rates that may result from broken or disconnected tubes.
We have also configured the alarms to be sensitive to a gas regulator outage, which is characterized by a complete loss of bubbling rate in a sector's exhaust bubblers, and may be accompanied by a nonzero bubbling rate in the associated relief bubblers.
The  timescale for triggering an alarm in response to a gas regulator outage is 30-50 minutes, corresponding to 3-5 bubble rate measurement intervals, and for triggering to broken tubes is 1-2 hours, corresponding to 6-12 bubble rate measurement intervals.

In the exhaust bubbler channels, where bubbles are expected during normal operation, an alarm is triggered if the bubbling rate falls outside a tunable range, defined by a specific number of standard deviations from the channel's mean bubbling rate. 
The mean and standard deviation used for setting alarm thresholds are determined from a period of normal gas flow. 
In the relief bubbler channels, where bubbles are not expected during normal operation, an alarm is triggered if any bubbles are observed.
The alarm configuration can be easily modified to be sensitive to other gas flow issues that may arise in the future.

Negligible damage is expected in the RPCs from gas flow interruptions on the  timescale of less than a week.
An alarm due to a low exhaust bubbling rate during beam operation will be addressed by checking for broken tubes or gas regulator problems on the next biweekly detector maintenance day.
For detector safety, the high voltage supplied to these RPCs is turned off if there is no gas flow through the RPCs.

\subsection{Monitoring GUI}
\label{subsec:gui}
\begin{figure*}[htb!]
    \centering
    \includegraphics[width=0.8\linewidth]{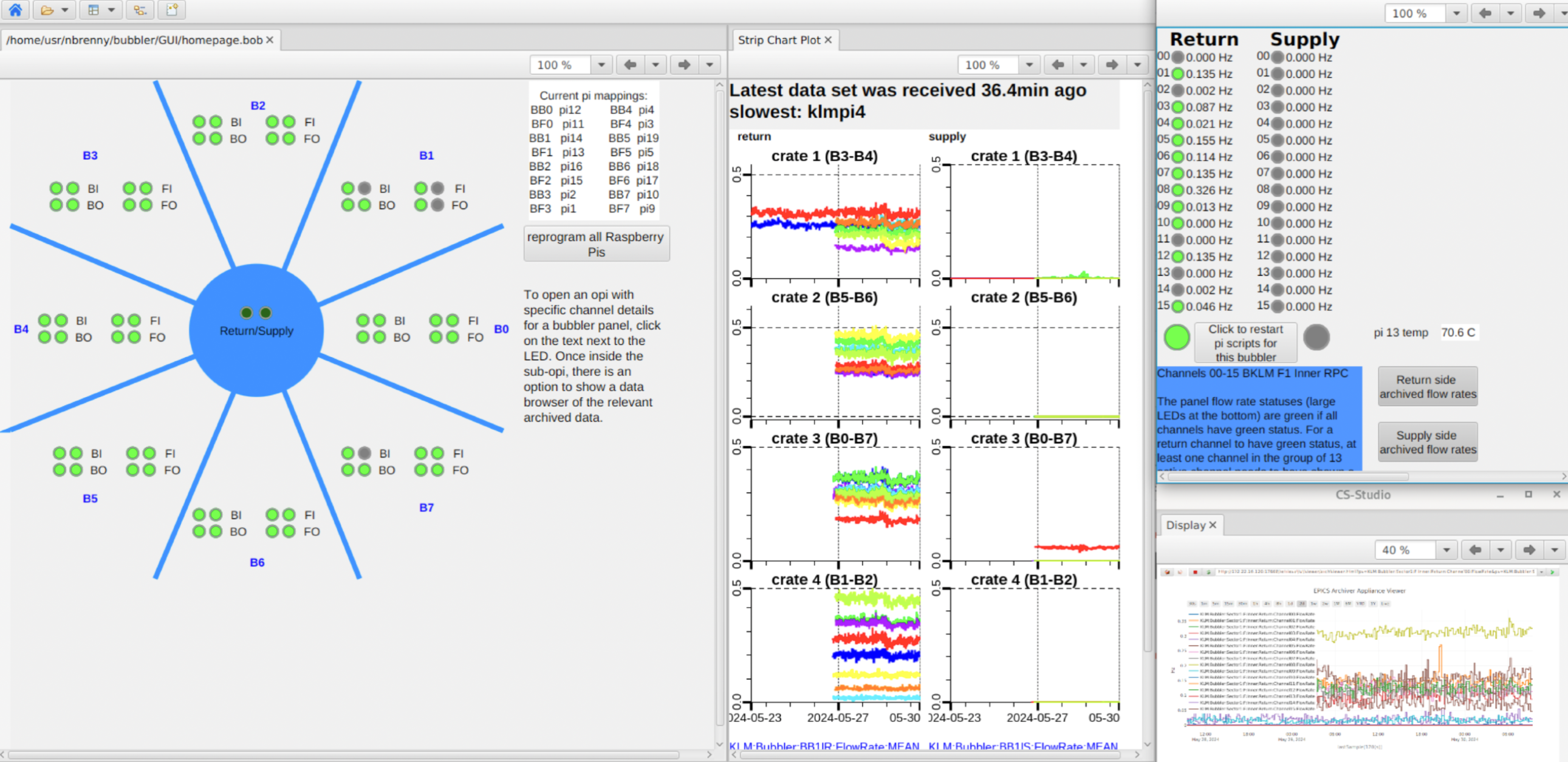}
    \caption{GUI for monitoring the bubbling rates. The left window displays indicators for the panels' bubbling rate statuses. The central window displays the means and standard deviations of the panels' bubbling rates. The upper right window displays the channel status for an individual panel. The lower right window displays bubble rate history.}
    \label{fig:gui}
\end{figure*}

To visualize the status of the RPC gas flow, we have designed a Graphical User Interface (GUI) (see Fig.~\ref{fig:gui}) to summarize the bubbling rates for each sector, which is available to the KLM operator during beam operations. 
The GUI is implemented using Control System Studio~\cite{CSS}.

It consists of a summary display with indicators for the bubbling rate statuses of the 64 bubbler panels, as well as more detailed displays for each bubbler panel.
Also displayed in the GUI are time series for the mean and standard deviation of the bubbling rates for each bubbler panel's 13 channels.
The mean and standard deviation are calculated every 10 minutes.
The KLM operator can check the bubble rate history in the GUI for more thorough investigations.
The GUI can also be used to reprogram the SBCs remotely.

\section{Performance}
\label{sec:peformance}
We have tested the performance of the alarm configuration using bubbling rate data from previous gas flow issues.
In December 2023, before the bubbler-monitoring system was completely commissioned, a gas regulator was accidentally unplugged.
It resulted in no bubbles in the exhaust bubblers and bubbles in the relief bubblers.
The alarm configuration correctly generated an alarm when presented with this failure-mode data, validating the configuration's sensitivity to  potential gas regulator outage problems. 

The bubbler-monitoring system was in use during the data-taking period from February 2024 to July 2024.
During a maintenance day in this period, four gas tubes at a bubbler panel located in a heavy-traffic location were damaged by inadvertent jostling.
The bubbler-monitoring system immediately discovered the decrease in flow.
Experts were then notified, and the tubes were replaced.
This situation demonstrated that the bubbler-monitoring system can detect interruptions in individual RPC gas flow rates.
No gas regulator problems were observed during the aforementioned data-taking period.

\section{Summary}
\label{sec:summary}
We have presented the hardware and software developments that constitute the integration of a new KLM RPC bubbler-monitoring system at \belletwo.
We have designed a readout and monitoring board consisting of analog multiplexers, ADCs, and an SBC to read analog bubble signals from the photosensors.
Software on board the SBC recognizes the bubble signals, computes bubbling rates, and broadcasts the bubbling rates to the \belletwo archiver.
The bubbling rates have been integrated into the \belletwo alarm system.
The integrated bubbler-monitoring system is capable of detecting a gas regulator issue like the one that contributed to a 50\% detection efficiency drop in one sector of the \belletwo KLM in 2021.
Furthermore, the bubbler-monitoring system is sensitive to gas flow outages of individual RPCs.
Broken gas supply tubes have been discovered by the bubbler-monitoring system, and no gas regulator problems have been observed during this system's commissioning operation in early 2024.

\section*{Acknowledgements}
This work was supported by
the U.S. Department of Energy and Research Award
No.DE-SC0021430 and via U.S. Belle II Operations administered by Brookhaven National Laboratory (DE-SC0012704). 

This acknowledgment is not to be interpreted as an endorsement of any statement made
by our institute, funding agency, government, or its representatives.

\bibliographystyle{apsrev4-1} 
\bibliography{references}

\end{document}